\documentclass[10pt,twocolumn,letterpaper]{article}

\usepackage[review]{cvpr}      

\usepackage{graphicx}
\usepackage{amsmath}
\usepackage{amssymb}
\usepackage{booktabs}
\usepackage[table,xcdraw]{xcolor}
\usepackage{multirow}
\usepackage{bbding}

\usepackage[pagebackref,breaklinks,colorlinks]{hyperref}

\usepackage[capitalize]{cleveref}
\crefname{section}{Sec.}{Secs.}
\Crefname{section}{Section}{Sections}
\Crefname{table}{Table}{Tables}
\crefname{table}{Tab.}{Tabs.}

\usepackage{enumitem}

\newcommand{\Mmat}[0]{{{\bf M}}}
\newcommand{\Nmat}[0]{{{\bf N}}}

\newcommand{\Qmat}[0]{{{\bf Q}}}

\newcommand{\Xmat}{{\bf X}}
\newcommand{\Ymat}[0]{{{\bf Y}}}

\newcommand{\iv}[0]{{\boldsymbol{i}}}

\newcommand{\nv}{\boldsymbol{n}}

\newcommand{\vv}{\boldsymbol{v}}

\newcommand{\xv}{\boldsymbol{x}}
\newcommand{\yv}{\boldsymbol{y}}

\newcommand{\Phimat}{\boldsymbol{\Phi}}

\DeclareMathOperator*{\argmin}{arg\,min}

\newcommand{\ts}{^{\top}}

\begin{document}
%
\title{Two-Stage is Enough: A Concise Deep Unfolding Reconstruction Network for Flexible Video Compressive Sensing}
%
%
%

\author{
\textbf{Siming Zheng\textsuperscript{1,2}, Xiaoyu Yang\textsuperscript{1,2}, Xin Yuan\textsuperscript{3}}\\
1.Computer Network Information Center, Chinese Academy of Science, Beijing, 100190, China\\
2.University of Chinese Academy of Sciences, Beijing 100049, China\\
3.School of Engineering, Westlake University, Hangzhou 310024, Zhejiang, China\\
{\tt\small zhengsiming@cnic.cn;} {\tt\small kxy@cnic.cn;} {\tt\small xyuan@westlake.ed.cn}
}

\date{}
\maketitle

\begin{abstract}
 We consider the reconstruction problem of video compressive sensing (VCS) under the deep unfolding/rolling structure. Yet, we aim to build a flexible and concise model using minimum stages. Different from existing deep unfolding networks used for inverse problems, where more stages are used for higher performance but without flexibility to different masks and scales, hereby we show that a 2-stage deep unfolding network can lead to the {state-of-the-art (SOTA)} results (with a 1.7dB gain in PSNR over the single stage model, RevSCI~\cite{cheng2021memory}) in VCS. The proposed method possesses the properties of adaptation to new masks and ready to scale to large data \textbf{without any additional training} thanks to the advantages of deep unfolding.
 Furthermore, we extend the proposed model for color VCS to perform joint reconstruction and demosaicing. Experimental results demonstrate that our 2-stage model has also achieved {SOTA} on color VCS reconstruction, leading to a $>$2.3dB gain in PSNR over the previous SOTA algorithm based on plug-and-play framework, meanwhile speeds up the reconstruction by $>$17 times.
 In addition, we have found that our network is also flexible to the mask modulation and scale size for color VCS reconstruction so that a single trained network can be applied to different hardware systems.  The code and models will be released to the public.
\end{abstract}

\setlength\parskip{.05\baselineskip}
\section{Introduction}
\label{sec:intro}
\begin{figure}[t!]
\centering
\includegraphics[width=1.0\columnwidth]{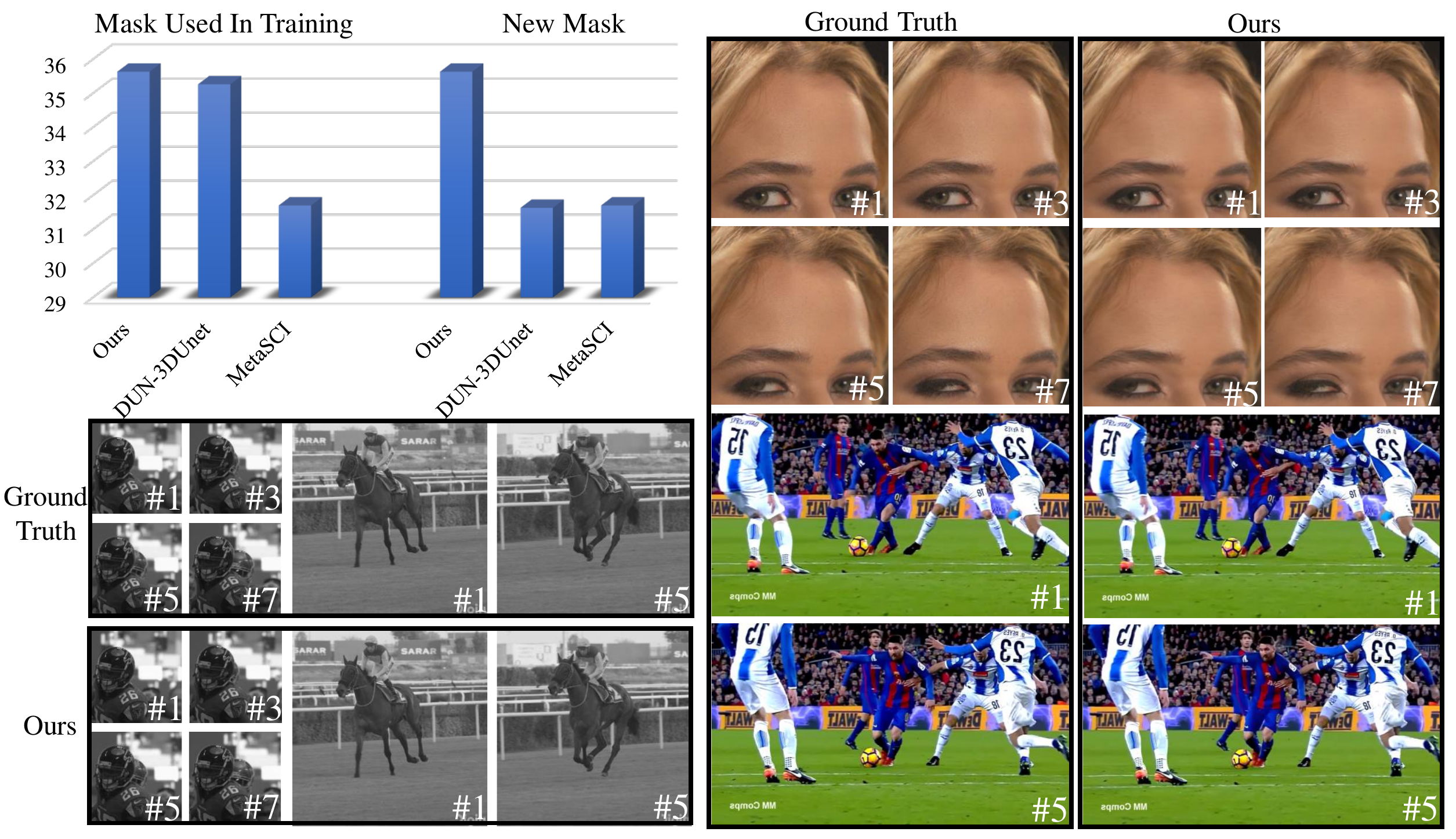}
\vspace{-5mm}
\caption{Our proposed concise deep unfolding model can not only provide state-of-the-art (SOTA) reconstruction (results shown in the bottom and top-right) for video compressive sensing but also be flexible to different modulations, \ie, providing stable results for different masks. In the top-left plot, we compare the average reconstruction PSNR on the benchmark simulation dataset using our proposed network with previous SOTA network, DUN-3DUnet~\cite{Wu_2021_ICCV} and MetaSCI~\cite{wang2021metasci} on both the pre-trained mask and new mask. It can be seen our method provides stable PSNR while DUN-3DUnet declines significantly on a new mask; even though MetaSCI keeps stable performance after adaptation, the reconstruction quality is not that satisfactory .}
\vspace{-2mm}
\label{fig:teasing_image}
\end{figure}

Video compressive sensing (VCS), a promising way to use a low-speed camera to capture high-speed scenes, employs the idea of compressive sensing~\cite{Candes06_Robust,Donoho06_CS} and implements the temporal compression using a high frequency modulation device, \eg, via a spatial light modulator~\cite{hitomi2011video} such as a digital micro-mirror device (DMD)~\cite{hitomi2011video,Reddy11_CVPR_P2C2} or a shifting mask~\cite{Llull13_OE_CACTI}, to modulate consequent frames and then integrates these modulated frames to a single {\em compressed measurement}. 
This single two-dimensional (2D) measurement can be used to reconstruct a high-speed video along with the modulation patterns using advanced algorithms~\cite{liu2018rank,cheng2021memory}. 
In a nutshell, VCS is composed of an optical encoder, \ie, the hardware imaging system, and a software decoder, \ie, the reconstruction algorithms~\cite{Yuan2021_SPM}. In this paper, we focus on the decoder part, and  specifically, using deep learning for the VCS reconstruction.

In general, VCS reconstruction belongs to the ill-posed inverse problem due to the fact that there are more unknown parameters (the desired video) to estimate than the known ones (the compressed measurement). 
Due to its efficiency and effectiveness, deep learning has recently been the dominant approach for VCS reconstruction. Towards this end, various deep models have been proposed:
\vspace{-2mm}
\begin{itemize}[leftmargin=*]
\setlength{\itemsep}{0pt}
\setlength{\parsep}{0pt}
\setlength{\parskip}{0pt}
    \item Single stage model takes the measurement (and optionally masks) as input and outputs the desired video directly.  U-net~\cite{qiao2020deep}, RevSCI~\cite{cheng2021memory} and MetaSCI~\cite{wang2021metasci} have been developed with different network structures. 
    \item Aiming to extract the temporal correlations among adjacent frames, recurrent neural networks have been used for VCS reconstruction, dubbed BIRNAT~\cite{cheng2020birnat}.
    \item As a novel approach to combine iterative algorithms and deep learning networks, deep unfolding/unrolling approaches~\cite{yang2016deep,yang2018admm} have also been used for VCS reconstruction ~\cite{meng2020gap,Wu_2021_ICCV,li2020end}. Another gain of deep unfolding is the interpretability~\cite{zhang2018ista}, where the deep networks play the role of denoising in each stage. 
\end{itemize}
\vspace{-2mm}

Most of the models are only suitable for the masks which are utilized for training. When the masks change but even with the same spatial size such as $256\times256$ or $512\times512$ pixels, these models can not achieve satisfactory results. Moreover, as the spatial size changes, the quality of the reconstruction results will also decline. For some of the models, training is limited due to the memory of GPU. Another line of work is based on the plug-and-play (PnP) framework~\cite{yuan2020plug,Yuan14CVPR}, which does not need re-train the model and is flexible to the mask modulation and scale size with a pre-trained deep image denoiser. Nevertheless, PnP based methods are still time consuming owing to the iterative nature.
Another method aiming for flexible reconstruction is using Meta learning~\cite{wang2021metasci}, unfortunately, the adaptation time is still long for new masks.
  
Combining the above considerations, in this paper, we design a {\bf flexible and concise two-stage deep unfolding network}, in which we fuse the {\em reference measurement frames} (RMF) into each stage. 3D convolutional neural network (CNN) based invertible structure is firstly utilized to save the memory during training under the deep unfolding framework which is always restricted on account of the model size. Furthermore, to address the limitation of flexibility, a {\em stage-by-stage training} strategy and {\em stage-wise loss} are introduced. The contributions of our work can be summarized as below:
\begin{itemize}
\setlength{\itemsep}{0pt}
    \item[1)] As presented in the literature, under the deep unfolding framework, the VCS reconstruction accuracy will increase with a higher stage/phase number. Considering the memory cost and inference time, we firstly introduce the {\em 3D-CNN based invertible structure} into the deep unfolding framework with {\bf only two stages} to break the bottleneck.
    \item[2)] To fully take advantage of the information from the measurement, we fuse the {\em Reference Measurement Frames} into each stage, which leads to a boost of 0.59dB on PSNR in the benchmark dataset.
    \item[3)] We introduce a {\em stage-by-stage training strategy and utilize stage-wise loss} to make the most of deep unfolding framework's strengths. This improves the model's fidelity and flexibility to different masks and scales. Additionally, we achieve state-of-the-art (SOTA) results on both grayscale and color VCS reconstruction and the models can adapt to different masks and scales without any additional training. Please refer to Fig.~\ref{fig:teasing_image} for the reconstruction results of our proposed network and also the flexibility of our method. 
\end{itemize}




\section{Related Work}


The essence of the VCS reconstruction is an ill-posed inverse problem. In recent years, numerous algorithms are developed. Nevertheless, the now available algorithms are still far from real applications. They are mainly confronted with three aspects of the problems, \ie, fidelity, inference speed, and flexibility. 

Numerous regularization items, \eg, total variant (TV)~\cite{yuan2016generalized} , deep denoiser ~\cite{yuan2020plug}, non-local low rank~\cite{liu2018rank}, Gaussian mixture model (GMM)~\cite{Yang14GMMonline}~\cite{Yang14GMM} are utilized as priors in the optimization-based methods. The preponderance of them is that they can be easily applied in different scenarios due to the flexibility to disparate masks and scales, but the fidelity is not satisfactory. In addition, the needed hundreds of iterations for the optimization lead to a long time reconstruction process. 
To overcome these shortcomings, deep learning is employed to conduct the whole reconstruction process. U-net~\cite{qiao2020deep} is the first End-to-End (E2E) deep learning method to solve the VCS reconstruction. The inference speed was significantly improved. BIRNAT~\cite{cheng2020birnat} is designed based on the bidirectional recurrent neural network (RNN) combining the attention map which can easily fit different compression ratios during training by virtue of RNN's structural advantages. RevSCI~\cite{cheng2021memory} achieves the optimum performance among the single-stage models, and its memory efficiency can led to high compression ratio and large scale reconstruction. These models all solve the problem by an E2E method but lack the interpretability. 
Naturally, combining the optimization-based and learning-based methods can bring together the merit of both sides. The development of deep unfolding approaches rises in response to the proper conditions. DUN-3DUnet~\cite{Wu_2021_ICCV}, which achieved the best results previously, employs 3D-Unet as the backbone with dense feature map fusion to break the limitation on the transmission of network information. 
Whereas, existing deep-unfolding based methods can not give full play to both sides' strengths. As the stage number which is crucial for higher reconstruction accuracy increases, the memory requirements will skyrocket, which limits the performance of the model. Beyond that, the flexibility is not so desirable when the masks and scales vary which is extremely common in real applications~\cite{zheng2021super}. 

Motivated by these above realistic requirements, MetaSCI~\cite{wang2021metasci} constructs a shared backbone for different systems with light-weight meta-modulation parameters, which can adapt quickly to new masks and be ready to scale to large data. Yet the fidelity is sacrificed, \ie, the PSNR is nearly 3.6dB lower than DUN-3DUnet and nearly 4dB lower than ours. Note that MetaSCI still need training (adaptation) when the model transfers to different systems. 

What we mentioned above is all about grayscale VCS. As for color VCS, it is worth nothing that the method achieving the optimum result is PnP-FastDVDnet~\cite{yuan2020plug} instead of any E2E neural network so far. RevSCI tried to recover a large color video scene, however due to the inflexibility to large masks, reconstruction can only be conducted piece by piece, which causes severe block artifacts.

Bearing all these concerns, we design a 2-stage deep unfolding network to address the issues mentioned above, which achieves SOTA results and is flexible on both grayscale and color VCS. Meanwhile, introducing the invertible structure improves the memory efficiency under the deep unfolding framework.


\begin{figure}[ht]
\centering
\vspace{-2mm}
\includegraphics[width=1.0\columnwidth]{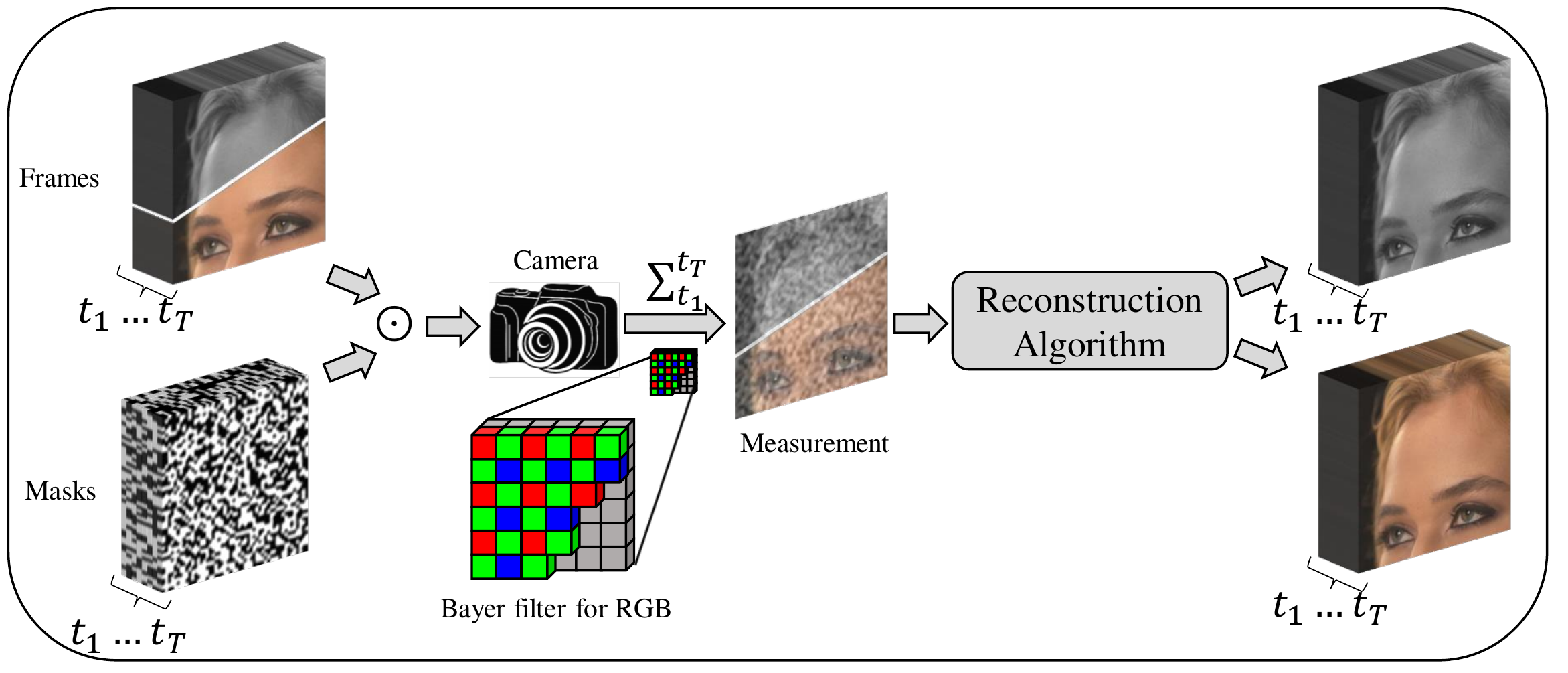}
\vspace{-5mm}
\caption{Pipeline of video compressive sensing, shown in grayscale in the top and RGB Bayer pattern in the bottom. $T$ consequent frames are modulated by $T$ different masks and then these modulated frames are summed up to a single measurement; reconstruction algorithms are then employed to reconstruct the high-speed high-quality video frames.}
\vspace{-5mm}
\label{fig:Pipline}
\end{figure}

\section{Mathematical Model of VCS}
Fig.~\ref{fig:Pipline} depicts the underlying principle of VCS, where multiple high-speed video frames (grayscale or color) are modulated by different masks and then these modulated frames are summed up to a single measurement. Our proposed concise network is then employed to reconstruct the high quality video frames either in grayscale or in color depending on the sensors being used. 

\subsection{Grayscale Video Compressive Sensing}
Let $\Xmat\in\mathbb{R}^{W\times H\times T}$ denote the to be captured video data-cube (frames) and $\Mmat\in\mathbb{R}^{W\times H\times T}$ denote the masks used to modulated each frame, where $W$, $H$, and $T$ denote the width, height, and the number of frames, respectively. The video frames modulated by the masks are $\Xmat'(:,:,t)=\Xmat(:,:,t)\odot \Mmat$, where $\Xmat'$ has the same size with $\Xmat$, $ t=1,2,\ldots,T $ and $\odot$ represents the Hadamard (element-wise) multiplication. Thus, the 2D VCS measurement $\Ymat\in\mathbb{R}^{W\times H}$ we get is the integral over time
across all the mask-modulated video frames, which can be represented by
\begin{align}
\textstyle \Ymat=\sum_{t=1}^{T}\Xmat'(:,:,t)+\Nmat,
\end{align}
where $\Nmat\in\mathbb{R}^\mathit{WH}$ denotes the measurement noise. By vectorizing the video frames and measurement, that is $\xv=\mathtt{vec}(\Xmat')\in\mathbb{R}^\mathit{WHB}$ and $\yv=\mathtt{vec}({\Ymat})\in\mathbb{R}^\mathit{WH}$, this model can be rewritten as
\begin{equation}
\label{eq:linear}
  \yv= \Phimat \xv+ \nv,
\end{equation}
where $\Phimat\in\mathbb{R}^{\mathit{WH}\times \mathit{WHB}}$ denotes the sensing matrix which is a concatenation of diagonal matrices, that is 
\begin{equation}
\label{eq:phi}
  \Phimat = [\Phimat_1,\dots,\Phimat_T],
\end{equation}
where $\Phimat_t= \mathtt{Diag}(\mathtt{vec}(\Mmat(:, :, t)))$ is the diagonal matrix with $\mathtt{vec}(\Mmat(:, :, t))$ as the diagonal elements. Note that $\Phimat$ is a very sparse matrix and the theoretical bounds of VCS have been developed in~\cite{jalali2019snapshot}.
\subsection{RGB Color Video Compressive Sensing}
Bayer filter mosaic is a color filter array (CFA) employed to arrange RGB filters on a square grid of light sensors, shown in the Fig.~\ref{fig:Pipline}. Its special color filter arrangement is utilized in most single-chip digital image sensors used in digital cameras, camcorders and scanners to create color images. As for the color VCS system, we bring the Bayer pattern filter sensor into consideration, which is half green (G), one quarter red (R) and one quarter blue (B), \ie, the RGGB pattern. The causation of utilizing more green channels is that the human retina uses both long wavelength-sensitive (L) and  middle wavelength-sensitive (M) cones to detect light during the day and is most sensitive to green light~\cite{otake2000spatial}. Let $\Xmat_{RGB}\in\mathbb{R}^{W\times H\times T\times 3}$ denote the to be captured RGB color video frames and $\Xmat_{RGGB}\in\mathbb{R}^{W\times H\times T}$ denote the captured video frames after the Bayer Filter. Under this situation, $\Xmat'(:,:,t)$ is a mosaicked frame which can be modeled as
\begin{equation}
  \Xmat'(:,:,t)=\Xmat_{RGGB}(:,:,t) \odot \Mmat(:,:,t).
\end{equation}

In most of the previous color VCS methods~\cite{Yuan14CVPR,yuan2021plug}, the original measurement was divided into four sub-measurements according the channels and these channels are reconstructed separately. Then these channels are interleaved and demosaiced to obtain the final color video. In our work we omit these steps and reconstruct the results directly and thus a joint framework is proposed.
\begin{figure*}[ht]
\centering
\includegraphics [width=1.8\columnwidth]{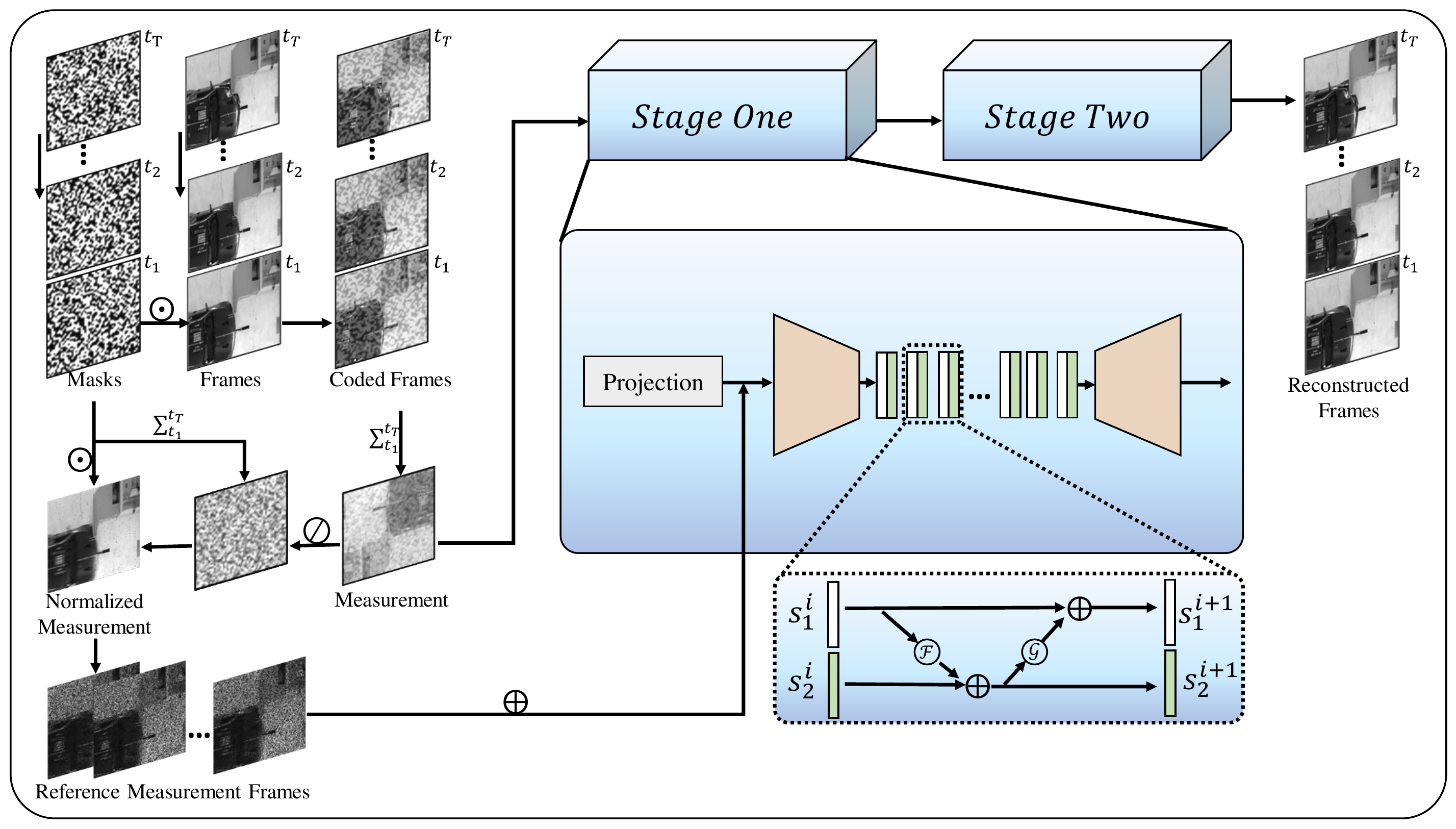}
\vspace{-3mm}
\caption{Illustration of our proposed model (grayscale VCS system). Left: the video frames are modulated by dynamic masks and then integrated along time to achieve the compressed measurement. By the Hadamard (element-wise) division between measurement and the summed mask, normalized measurement is obtained. After the element-wise multiplication between the normalized measurement and masks, the Reference Measurement Frames (RMF) are obtained. Middle: the network's architecture in each stage. The backbone is an encoder-decoder structure with {\em invertible blocks} plugged in and the RMFs are fed into each stage. Right: reconstructed video frames. } 
\label{fig:Network}
\vspace{-3mm}
\end{figure*}

\section{Proposed Method}
In this section, we describe the proposed network in detail. Our concise model only consists of two stages, but novel techniques of reference measurement frames, invertible structure and stage-wise training are significantly important to provide good results. 
In the following, we first describe the network based on the grayscale VCS and extend it to the color case where a joint reconstruction and demosaicing model is introduced. 
\subsection{Model for Grayscale VCS}
\noindent{\bf Reference Measurement Frames:}
Inspired by \cite{cheng2020birnat,cheng2021memory}, we introduce the RMF into our network and fuse the output of each stage's projection and the RMF together. As shown in the left part of Fig.~\ref{fig:Network}, the normalized measurement $\overline{\Ymat}$ can be expressed as
\begin{equation}
  \overline{\Ymat}=\Ymat\oslash\Mmat',
\end{equation}
where $\oslash$ is the Hadamard (element-wise) division and $\Mmat'=\sum_{t=1}^{T}\Mmat(:,:,t)$ is the integral of masks across time,  where each element in $\Mmat'$ describes how many corresponding pixels of $\Xmat$ are integrated into the measurement $\Ymat$. If we directly concatenate $\Ymat$ into the network, the dynamic range gap between $\Ymat$ and the desired result makes it less effective to train. It can be easily found that $\overline{\Ymat}$ contains more visual information than $\Ymat$. The regions with little motion such as backgrounds resemble the original scenes. Furthermore, we concatenate $\hat{\Xmat}=\overline{\Ymat}\odot\Mmat$, which could be used as a reference of the modulated frames, with the output of each stage's projection (introduced below). 

\noindent{\bf Deep Unfolding Network For VCS:}
The VCS reconstruction problem can be considered to solve the following optimization problem:
\begin{equation}
\label{eq:oriproblem}
  \hat{\xv} = \textstyle \argmin_{\xv} \|\yv - \Phimat \xv\|_2^2 + \tau R(\xv), 
\end{equation}
where $R(\xv)$ denotes the prior term to regularize the inverse problem and $\tau$ denotes the regularization parameter. To solve Eq.~\ref{eq:oriproblem}, conventional optimization-based methods divided it into sub-problems according to the optimization algorithms such as TwIST~\cite{Bioucas-Dias2007TwIST}, GAP-TV~\cite{yuan2016generalized} and solve it by iterative computations. We employ the framework in \cite{meng2020gap}, which combines the advantages of both deep unfolding and generalized alternating projection~\cite{Liao14GAP} ideas and is also tightly connected to the plug-and-play (PnP) framework~\cite{yuan2020plug}.
Each stage in our network is composed of the following two steps:
\begin{align}
     & \xv^{(j)}=\vv^{(j-1)} + \Phimat\ts(\Phimat\Phimat\ts)^{-1}(\yv - \Phimat  \vv^{(j-1)}), \label{eq:pnpGAP1}\\
     & \vv^{(j)}= {\rm Network}([\xv^{(j)},\Xmat_R]_4), \label{Eq:pnpGAP2}
\end{align}
where $\vv^{(j)}$ denotes our current estimate of the desired signal at the $j$-th stage, $\Xmat_R$ denotes the RMF, {$[~]_4$ denotes the concatenation across the fourth dimension, \ie, the filter channel of 3D-CNN}. From Eq.~\eqref{eq:phi}, we have 
    $\Qmat \stackrel{\rm def}{=} \Phimat\Phimat\ts = \mathtt{Diag}(Q_1,\dots,Q_{WH})$
is a diagonal matrix where $Q_i=\sum_{t=1}^{T}\Mmat_{t,i}^2$ , $\forall i=1,\dots,WH$. Thus Eq.~\eqref{eq:pnpGAP1}, a.k.a., the projection step can be computed efficiently. After the projection, a deep neural network is utilized to regularize the output of Eq.~\eqref{eq:pnpGAP1} with the help of $\Xmat_R$ which has more visual information about the motionless areas such as the edges and backgrounds. In PnP framework, the prior regularization is a flexible deep denoiser and the whole reconstruction process consists of hundreds of iterations which are time consuming. Despite the results achieved by utilizing a denoiser as the prior regularization is not too bad, we are still not aware of where is the exactly desired signal domain. Hence we use the E2E network to get as close to the optimal domain as possible. The related proof of the convergence can be found in \cite{meng2020gap}.

Our backbone shares the resembling encoder-decoder structure with invertible blocks plugged in as designed in \cite{cheng2021memory}. Whereas we extend the input channel number to take the information of the RMF into consideration, which relies on the 3D-CNN's advantage that it is adept at capturing the  information of spatial-temporal correlation. Besides we increase the number of channels in the model. Furthermore, we reduce the number of the invertible blocks nearly by half for the sake of the model's concision.

As shown in the bottom right of Fig.~\ref{fig:Network}, the extracted feature map is fed into the invertible blocks. We follow the original setting in \cite{gomez2017reversible} that we split the input feature map into two parts by channel. The forward pass can be modeled as
\begin{equation}
\begin{split}
& s_1^{\iv+1} =s_1^{\iv}+\mathcal{F}(s_2^{\iv}), ~~~s_2^{\iv+1} =s_2^{\iv}+\mathcal{G}(s_1^{\iv+1}).
\end{split}
\end{equation}
Then the inverse pass can be directly computed by the following with known $s_1^{\iv+1}$ and $s_2^{\iv+1}$:
\begin{equation}
\begin{split}
  & s_2^{\iv} =s_2^{\iv+1}-\mathcal{G}(s_1^{\iv+1}), ~~~s_1^{\iv} =s_1^{\iv+1}-\mathcal{F}(s_2^{\iv}).
\end{split}
\label{eq:back}
\end{equation}
After the propagation through the invertible blocks, the feature map is integrated to obtain the desired result of the current stage. All the settings are the same for the two stages in our model.

\subsection{Model for Color VCS}
Our proposed model is the first deep network that outperforms the PnP-FastDVDnet~\cite{yuan2021plug} on the color VCS task. Note that PnP is still an iterative algorithm and thus it is realtively slow in real applications.
Compared to the model of grayscale VCS, the model of color VCS shares almost the same architecture. However, due to the captured measurement of color VCS system is a Bayer mosaicked measurement, we do not fuse the RMF into the model to avoid the unknown adverse factors. Different from previous methods such as \cite{cheng2021memory,Yuan14CVPR} which divide the measurement into four individual parts corresponding to the Bayer-filter, our proposed model directly takes the whole Bayer measurement as input and outputs the desired color videos at each stage. In addition, the clean output of the first stage is captured by the Bayer filter again, then we feed it into the next stage's projection. In this manner, we achieve the high performance on color VCS systems.

\begin{table*}[ht]
\caption{ The quantitative comparison of different algorithms for grayscale VCS system. The average results of PSNR in dB (left entry), SSIM (right entry) and running time per measurement in seconds. Note that GAP-TV and DeSCI are running on CPU while others are on GPU. The best results are bold, and the second best results are \underline{underlined}. The results of our model with three stages can be found in SM.}
\vspace{-3mm}
\centering
	\resizebox{.96\textwidth}{!}
	{
\begin{tabular}{
>{\columncolor[HTML]{FFFFFF}}c 
>{\columncolor[HTML]{FFFFFF}}c 
>{\columncolor[HTML]{FFFFFF}}c 
>{\columncolor[HTML]{FFFFFF}}c 
>{\columncolor[HTML]{FFFFFF}}c 
>{\columncolor[HTML]{FFFFFF}}c 
>{\columncolor[HTML]{FFFFFF}}c 
>{\columncolor[HTML]{FFFFFF}}c 
>{\columncolor[HTML]{FFFFFF}}c }
\hline
{\color[HTML]{000000} Dataset}        & {\color[HTML]{000000} Kobe}                 & {\color[HTML]{000000} Traffic}     & {\color[HTML]{000000} Runner}               & {\color[HTML]{000000} Drop}        & {\color[HTML]{000000} Aerial}               & {\color[HTML]{000000} Crash}                & {\color[HTML]{000000} Average}              & {\color[HTML]{000000} Running time} \\ \hline
{\color[HTML]{000000} GAP-TV}         & {\color[HTML]{000000} 26.45 0.845}          & {\color[HTML]{000000} 20.90 0.715} & {\color[HTML]{000000} 28.48 0.899}          & {\color[HTML]{000000} 33.81 0.963} & {\color[HTML]{000000} 25.03 0.828}          & {\color[HTML]{000000} 24.82 0.838}          & {\color[HTML]{000000} 26.58 0.848}          & {\color[HTML]{000000} 4.2}          \\
{\color[HTML]{000000} E2E-CNN}        & {\color[HTML]{000000} 27.79 0.807}          & {\color[HTML]{000000} 24.62 0.840} & {\color[HTML]{000000} 34.12 0.947}          & {\color[HTML]{000000} 36.56 0.949} & {\color[HTML]{000000} 27.18 0.869}          & {\color[HTML]{000000} 26.43 0.882}          & {\color[HTML]{000000} 29.45 0.882}          & {\color[HTML]{000000} 0.0312}       \\
{\color[HTML]{000000} DeSCI}          & {\color[HTML]{000000} 33.25 0.952}          & {\color[HTML]{000000} 28.72 0.925} & {\color[HTML]{000000} 38.76 0.969}          & {\color[HTML]{000000} 43.22 0.993} & {\color[HTML]{000000} 25.33 0.860}          & {\color[HTML]{000000} 27.04 0.909}          & {\color[HTML]{000000} 32.72 0.935}          & {\color[HTML]{000000} 6180}         \\
{\color[HTML]{000000} PnP-FFDNet}     & {\color[HTML]{000000} 30.47 0.926}          & {\color[HTML]{000000} 24.08 0.833} & {\color[HTML]{000000} 32.88 0.938}          & {\color[HTML]{000000} 40.87 0.988} & {\color[HTML]{000000} 24.020.814}           & {\color[HTML]{000000} 24.32 0.836}          & {\color[HTML]{000000} 29.44 0.889}          & {\color[HTML]{000000} 3.0}          \\
{\color[HTML]{000000} PnP-FastDVDNet} & {\color[HTML]{000000} 32.73 0.946}          & {\color[HTML]{000000} 27.95 0.932} & {\color[HTML]{000000} 36.29 0.962}          & {\color[HTML]{000000} 41.82 0.989} & {\color[HTML]{000000} 27.98 0.897}          & {\color[HTML]{000000} 27.32 0.925}          & {\color[HTML]{000000} 32.35 0.942}          & {\color[HTML]{000000} 18}           \\
{\color[HTML]{000000} BIRNAT}         & {\color[HTML]{000000} 32.71 0.950}          & {\color[HTML]{000000} 29.33 0.942} & {\color[HTML]{000000} 38.70 0.976}          & {\color[HTML]{000000} 42.28 0.992} & {\color[HTML]{000000} 28.99 0.927}          & {\color[HTML]{000000} 27.84 0.927}          & {\color[HTML]{000000} 33.31 0.951}          & {\color[HTML]{000000} 0.16}         \\
{\color[HTML]{000000} GAP-Unet-S12}   & {\color[HTML]{000000} 32.09 0.944}          & {\color[HTML]{000000} 28.19 0.929} & {\color[HTML]{000000} 38.12 0.975}          & {\color[HTML]{000000} 42.02 0.992} & {\color[HTML]{000000} 28.88 0.914}          & {\color[HTML]{000000} 27.83 0.931}          & {\color[HTML]{000000} 32.86 0.947}          & {\color[HTML]{000000} 0.0072}       \\
{\color[HTML]{000000} MetaSCI}        & {\color[HTML]{000000} 30.12 0.907}          & {\color[HTML]{000000} 26.95 0.888} & {\color[HTML]{000000} 37.02 0.967}          & {\color[HTML]{000000} 40.61 0.985} & {\color[HTML]{000000} 28.31 0.904}          & {\color[HTML]{000000} 27.33 0.906}          & {\color[HTML]{000000} 31.72 0.926}          & {\color[HTML]{000000} 0.025}        \\
{\color[HTML]{000000} RevSCI}         & {\color[HTML]{000000} 33.72 0.957}          & {\color[HTML]{000000} 30.02 0.949} & {\color[HTML]{000000} 39.40 0.977}          & {\color[HTML]{000000} 42.93 0.992} & {\color[HTML]{000000} 29.35 0.924}          & {\color[HTML]{000000} 28.12 0.937}          & {\color[HTML]{000000} 33.92 0.956}          & {\color[HTML]{000000} 0.19}         \\
{\color[HTML]{000000} DUN-3DUnet}     & {\color[HTML]{000000} \underline{35.00} \underline{0.969}}          & {\color[HTML]{000000} \textbf{31.76} \underline{0.966}} & {\color[HTML]{000000} \underline{40.90} \underline{0.983}}          & {\color[HTML]{000000} \textbf{44.46} \underline{0.994}} & {\color[HTML]{000000} \underline{30.46} \underline{0.943}}          & {\color[HTML]{000000} \underline{29.35} \underline{0.955}}          & {\color[HTML]{000000} \underline{35.32} \underline{0.968}}          & {\color[HTML]{000000} 1.35}         \\ \hline
{\color[HTML]{000000} Ours}           & {\color[HTML]{000000} \textbf{35.24} \textbf{0.984}} & {\color[HTML]{000000} \underline{31.45} \textbf{0.977}} & {\color[HTML]{000000} \textbf{41.47} \textbf{0.994}} & {\color[HTML]{000000} \underline{44.43} \textbf{0.997}} & {\color[HTML]{000000} \textbf{30.86} \textbf{0.965}} & {\color[HTML]{000000} \textbf{30.32} \textbf{0.976}} & {\color[HTML]{000000} \textbf{35.63} \textbf{0.982}} & {\color[HTML]{000000} 1.43}\\    \hline
\end{tabular}}
\label{table:gray}
\vspace{-5mm}
\end{table*}
\subsection{Stage-By-Stage Training and Stage-Wise Loss}
Given the training pairs $\{\yv_i,\xv_i\}_{\iv=1}^{k}$ where ${k}$ is the training sample number, $\yv_i$ is the measurement and $\xv_i$ is the ground truth. We propose a {\em stage-by-stage} training strategy.  Firstly, we only train the first stage. We chose the mean square error (MSE) as our loss function, \ie,
\begin{equation}
   \textstyle  \mathcal{L}_{MSE}^{(1)} =\frac{1}{cBTWH}\sum_{i=1}^{B}\|\hat{\xv}_i^{(1)}-{\xv}_i\|_2^2,
\end{equation}
where $B$ denotes the training batch size, the superscript `(1)' means the stage one, $\hat{\xv}_i^{(1)}$ denotes the reconstruction result of the first stage and $c$ is the channel of $\hat{\xv}_i^{(1)}$, \ie, $c=1$ for grayscale frames and $c=3$ for color frames. After training the first stage we fix the parameters of it. Then we add and train the second stage with the same setting as stage one. 

Recalling the previous work~\cite{metzler2017learned}, in which the denoising-based approximate message passing (LDAMP) is built on the D-AMP algorithm~\cite{metzler2016denoising} and it tried a layer-by-layer training method for a 10 layer network. Different from that, after we separately trained each stage, we unfix all the parameters and {\em refine the whole network} in an E2E manner which is inspired by the stochastic gradient descent theory which suggests that the stage-by-stage (layer-by-layer) should sacrifice performance as compared to E2E training~\cite{smieja1993neural}. When we train the network by the E2E training way, the loss we use is a stage-wise loss, \ie,
\begin{equation}
    \textstyle   \mathcal{LOSS} = 0.5  \mathcal{L}_{MSE}^{(1)}+ \mathcal{L}_{MSE}^{(2)},
\end{equation}
where the superscript `{(2)}' denotes the stage two. {The weight for different stages' loss is to approximate the first stage's gradient range during the stage-by-stage training.}

As for the process of the back-propagation, it is different with the normal way. Thanks to the invertible structure, the activations produced from the forward pass of the invertible blocks are not needed to be stored. During the training we only need to save the last block's activations and the previous blocks' activations can be calculated by Eq.~\eqref{eq:back}, the other parts of the network are trained as usual. 

\begin{figure}[!htbp]
\centering
\vspace{-2mm} 
\includegraphics[width=1.0\columnwidth]{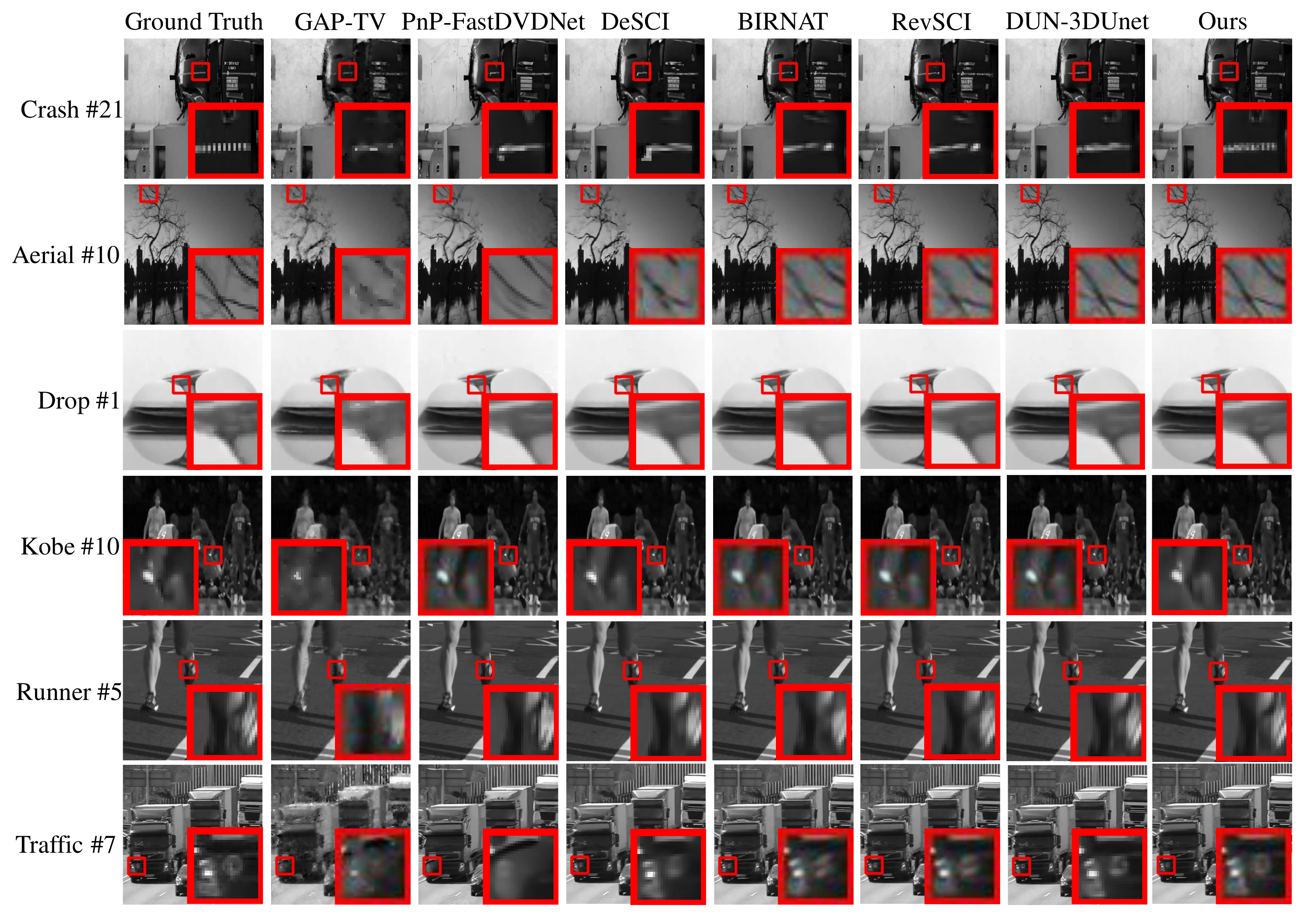}
\vspace{-5mm} 
\caption{Comparison of selected grayscale reconstruction video frames of six grayscale benchmark datasets ($256\times256\times8$).}
\label{fig:Gray simulation}
\vspace{-4mm} 
\end{figure}

We train the network on 2 NVIDIA A40 GPUs utilizing PyTorch~\cite{paszke2019pytorch}. Note that when we use the stage-wise loss, the calculated gradients of the repetitive part of the network are accumulated. The graph need to be retained when we utilize its automatic differentiation routine. 
The initial learning rate is set to $5\times{10}^{-5}$ for the first 8 epochs and  decays with a factor of 0.5 every 5 epochs.  Adam~\cite{kingma2014adam} is employed as the optimizer.


\section{Experiments}
\textbf{DAVIS 2017}~\cite{pont20172017} is chosen as our training dataset which has 90 different scenes with two resolutions: 480P and 1080P. The grayscale testing datasets follow \cite{liu2018rank} including \textbf{Kobe, Traffic, Runner, Drop, Crash,} and \textbf{Aerial} with the size of $256\times256\times8$. We randomly crop 27440 samples with the same size from the data of 480P and augment the data with rotations and flips. The color testing dataset is generated by \cite{yuan2020plug} including \textbf{Beauty, Bosphorus, Jockey, Runner, ShakeNDry,} and \textbf{Traffic} with the size of $512\times512\times3\times8$. The setting of color training dataset is basically the same with grayscale except that the spatial size changes to $256\times256\times3\times8$ and the resolution we choose is 1080P for more details. Note that due to the flexibility of our model, even the testing spatial size is $512\times512$, we still utilize data of $256\times256$ to train the model for efficiency. As for the large scale data, we follow \cite{wang2021metasci} which is also focused on the flexibility of VCS systems. For all the simulation testing data, both PSNR and structured similarity (SSIM)~\cite{wang2004image} are employed to evaluate the reconstruction quality.
\subsection{Simulation Results}

\begin{table*}[!htbp]
\caption{The quantitative comparison of different algorithms for color VCS system. The average results of PSNR in dB (left entry), SSIM (right entry) and running time per measurement in seconds. The best results are bold, and the second best results are \underline{underlined}.}
\vspace{-3mm}
\centering
	\resizebox{.9\textwidth}{!}
	{
\begin{tabular}{ccccccccc}
\hline
Dataset       & Beauty               & Bosphorus            & Jockey               & Runner               & ShakeNDry            & Traffic              & Average               & Running time \\ \hline
GAP-TV          & 33.08 0.964          & 29.70 0.914          & 29.48 0.887          & 29.10 0.878          & 29.59 0.893          & 19.84 0.645          & 28.47 0.864           & 10.8     \\
DeSCI & 34.66 0.971          & 32.88 0.952          & 34.14 0.938          & 36.16 0.949         & 30.94 0.905          & 24.62 0.839          & 32.23 0.926           & 92640    \\
PnP-FastDVDnet  & \underline{35.12} \underline{0.971}          & \underline{35.80} \underline{0.970}          & \underline{35.14} \underline{0.944}          & \underline{38.00} \underline{0.962}          & \underline{33.35} \underline{0.945}          & \underline{27.22} \underline{0.909}           & \underline{34.11} \underline{0.950}            & 52.2     \\ \hline
Ours            & \textbf{37.18 0.988} & \textbf{38.44 0.988} & \textbf{37.89 0.982} & \textbf{41.19 0.992} & \textbf{35.17 0.971} & \textbf{28.73 0.950} & \textbf{36.43 0.9785} & 3.06      \\ \hline
\end{tabular}}
\vspace{-3mm}
\label{table:RGB}
\end{table*}

\noindent{\bf{Grayscale Data}:}
We compare our model with 4 optimization-based (including PnP) methods, \ie, GAP-TV~\cite{yuan2016generalized}, PnP-FFDnet~\cite{yuan2020plug}, PnP-FastDVDnet~\cite{Yuan14CVPR} and DeSCI~\cite{liu2018rank}, and six learning-based methods including E2E-CNN~\cite{qiao2020deep}, GAP-Unet-S12~\cite{meng2020gap}, BIRNAT~\cite{cheng2020birnat}, MetaSCI~\cite{wang2021metasci}, RevSCI~\cite{cheng2021memory} and DUN-3DUnet~\cite{Wu_2021_ICCV}. The existing best algorithm is DUN-3DUnet. The quantitative comparison is shown in Table.~\ref{table:gray}. Selected reconstructed frames are shown in Fig.~\ref{fig:Gray simulation}.
It can be observed that our proposed method (with two stages) outperforms all the other algorithms in SSIM and surpass the available best algorithm DUN-3DUnet by 0.31dB in PSNR on average. 
As shown in Fig.~\ref{fig:Gray simulation}, optimization-based methods, \ie, GAP-TV and DeSCI achieve low fidelity results; PnP-FastDVDnet's results are over smooth. Learning-based methods such as BIRNAT and RevSCI achieve fairly good results but still miss details and it is a little blurry around the edges. Compared with the existing best algorithm DUN-3DUnet, our method can obtain more detailed information in the areas with large motions.
We do notice that more stages are able to lead better but limited gains (0.3dB improvement in PSNR for 3 stages in SM). However, it is not deserved to do this counting the increased parameters and training time. Therefore, we focus on the minimum, \ie, 2, stage model in this paper.

\begin{figure}[!htbp]
\centering
\vspace{-3mm}
\includegraphics[width=1.0\columnwidth]{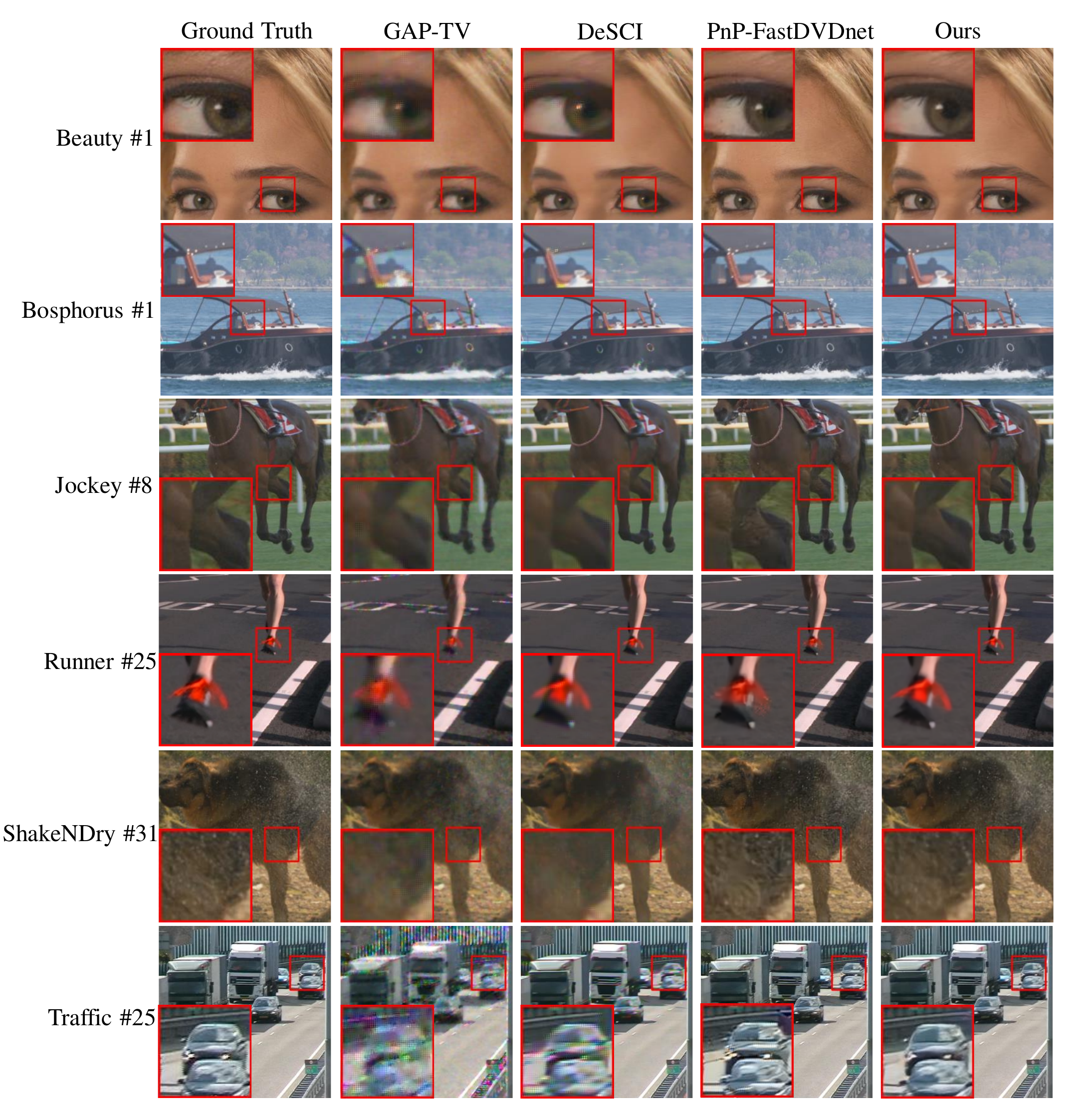}
\vspace{-8mm}
\caption{Comparison of selected color reconstruction video frames of color benchmark datasets ($512\times512\times3\times8$).}
\label{fig:color simulation}
\vspace{-5mm}
\end{figure}

\noindent{\bf{Color Data}:}
The existing best algorithm for color VCS is  PnP-FastDVDNet. We compare our results with GAP-TV, DeSCI, and PnP-FastDVDNet in Fig.~\ref{fig:color simulation} with quantitative comparison in Table.~\ref{table:RGB}. It can be noticed that the reconstructed images of GAP-TV are neither clean nor clear. DeSCI is kind of over smooth and there are artifacts in the regions with large motions which can be seen in the sample of `Traffic'. In the zooming areas, we can see that PnP-FastDVDNet leads to blurry results. Our method achieves the best reconstruction results with  sharp edges. We want to emphasize that the reconstructed color results with spatial size of $512\times512$ are obtained by the model trained on the data of spatial size of $256\times256$, which also reflects the flexibility of our proposed model.

\subsection{Flexibility}
\noindent{\bf{Different Masks}:}
To test the model's flexibility to masks, we randomly generate three masks that were not used in training. We compare our proposed model with another two algorithms. The reason we choose these two methods is that DUN-3DUnet is also based on the deep unfolding framework and is the existing best algorithm, and MetaSCI is designed specifically for the model adaptation. The average results of the models with different masks fed in are shown in Table.~\ref{table:different mask}. The testing data is the same with those used in grayscale VCS. It can be noticed that DUN-3DUnet's result quality degrades nearly 3.5dB and it barely can be flexible with low fidelity requirements for reconstruction. MeatSCI sacrifices the fidelity to improve the speed of adaptation, but the result quality of the model, which utilizes the mask used in training, is even worse than DUN-3DUnet's when DUN-3DUnet utilizes the masks which are not used in training.  Besides, MeataSCI still needs additional training (adaptation) for new masks. Note that the quality of our model's reconstruction result does not decline when using different masks which is not utilized in training. The result's quality may even improve a little bit when the mask changes.
\begin{table}[!htbp]
\vspace{-3mm}
\caption{Mask flexibility: quantitative comparison of different algorithms with different masks for grayscale VCS system.}
\vspace{-3mm}
\centering
	\resizebox{.45\textwidth}{!}
	{
\begin{tabular}{|c|c|c|c|}
\hline
Algorithm           & \textbf{Ours}        & DUN-3DUnet  & MetaSCI     \\ \hline
Mask Used In Training & \textbf{35.63 0.982} & 35.26 0.968 & 31.70 0.925 \\ \hline
New Mask\_1         & \textbf{35.64 0.982} & 31.72 0.937 & NA          \\ \hline
New Mask\_2         & \textbf{35.61 0.982} & 31.63 0.937 & NA          \\ \hline
New Mask\_3         & \textbf{35.63 0.982} & 31.74 0.937 & NA          \\ \hline
\end{tabular}}
\label{table:different mask}
\vspace{-3mm}
\end{table}

\noindent{\bf{Different Scales}:} Few researchers focus on large scale data for VCS no matter in grayscale or color systems. Thanks to the flexibility, some conventional optimization-based algorithms such as PnP-FFDNet~\cite{yuan2020plug} and GAP-TV~\cite{yuan2016generalized} can adapt to different scales both in grayscale and color VCS systems. Of all the learning-based algorithms in existence, only MetaSCI~\cite{wang2021metasci} can be used for grayscale large scale reconstruction of VCS. We utilize the same benchmark following \cite{wang2021metasci} and test our model with quantitative comparisons presented in the Table.~\ref{table:scale result} and selected reconstructed frames shown in the supplementary material (SM). It can be observed that our proposed model outperforms all the existing methods with a huge gap of nearly 5dB comparing with the existing best method. Note that we do not retrain the model which is trained by the data with the spatial size of $256\times256$. Due to the limitation of the GPU memory, we divide the data with the size of $2048\times2048$ into four blocks and reconstruct the results block by block relying on the model's flexibility. More results are shown in the SM.

\setlength{\parskip}{0.1cm plus1mm minus5mm}
\subsection{Ablation Study} 
To quantitatively verify the effectiveness of the three parts, \ie, RMF, stage-by-stage training strategy and stage-wise loss in our model, we conduct comparative experiments on the grayscale benchmark dataset with results shown in Table.~\ref{table:Ablation Study}. It can be seen that without the fusion of RMF, the PSNR decreases by nearly 0.6dB. But the model is still with decent flexibility. Then we train the model by an E2E way from the very beginning and the loss we use is the stage-wise loss (with RMF fused in), the PSNR decreases by nearly 0.4dB and the model is also flexible. Next we only change the training loss just utilizing $\mathcal{L}_{MSE}^{(2)}$ instead, which brings a severe drop in both PSNR and flexibility. The drop of the PSNR is nearly 1.2dB, and after changing the mask the PSNR gets another decline of 0.4dB. Finally we test the model (with RMF fused in) trained by stage-by-stage strategy but without refining after training the two stages separately, the PSNR only decreases 0.2dB but decreases another 0.5dB after changing the mask. In summary, the stage-by-stage training strategy and stage-wise loss contribute to the model's flexibility significantly by fully training each stage to get as close to the optimum desired signal domain as possible after each projection; and all the three parts contribute to the final results. 
\begin{table*}[h]
\caption{Scale flexibility: quantitative comparison (average PSNR, SSIM and running time) of different algorithms for grayscale VCS system under different scales, best results in {\bf bold}, and the second best \underline{underlined}. The results of $512\times512$ are in SM.}
\vspace{-3mm}
\centering
	\resizebox{.83\textwidth}{!}
	{
\begin{tabular}{ccccccccc}
\hline
Size                       & Algorithm  & Beauty               & Jockey               & ShakeNDry            & ReadyGo              & YachtRide            & Average              & Test Time \\ \hline
\multirow{4}{*}{1024x1024} & GAP-TV     & 33.59 0.852          & 33,27 0.971          & 33.86 0.913          & 27.49 0.948          & 24.39 0.937          & 30.52 0.924          & 178.11    \\
                           & PnP-FFDNet & 32.36 0.857          & 35.25 0.976          & 32.21 0.902          & 31.87 0.965          & 30.77 \underline{0.967}          & 32.49 0.933          & 52.47     \\
                           & MetaSCI    & \underline{35.23} \underline{0.929}          & \underline{37.15} \underline{0.978}          & \underline{36.06} \underline{0.939}          & \underline{33,34} \underline{0.973}          & \underline{32.68} 0.955          & \underline{34.89} \underline{0.95}5          & 0.59      \\
                           & Ours       & \textbf{40.34 0.979} & \textbf{42.25 0.989} & \textbf{38.73 0.976} & \textbf{40.13 0.988} & \textbf{37.38 0.981} & \textbf{39.77 0.983} & 14.42     \\ \hline
Size                       & Algorithm  & City                 & Kids                 & Lips                 & Niqht                & RiverBank            & Average              & Test Time \\ \hline
\multirow{4}{*}{2048x2048} & GAP-TV     & 21.27 0.902          & 26.05 0.956          & 26.46 0.890          & 26.81 0.875          & 27.74 0.848          & 25.67 0.894          & 764.75    \\
                           & PnP-FFDNet & 29.31 0.926          & 30.01 \underline{0.966}          & 27.99 \underline{0.902}          & 31.18 0.891          & 30.38 0.888          & 29.77 0.915          & 205.62    \\
                           & MetaSCI    & \underline{32.63} \underline{0.930}          & \underline{32.31} 0.965          & \underline{30.90} 0.895         & \underline{33.86} \underline{0.893}          & \underline{32.77} \underline{0.902}          & \underline{32.49} \underline{0.917}          & 2.38      \\
                           & Ours       & \textbf{40.40 0.983} & \textbf{40.13 0.984} & \textbf{35.50 0.933} & \textbf{36.47 0.955} & \textbf{36.82 0.970} & \textbf{37.86 0.965} & 57.71      \\ \hline
\end{tabular}}
\label{table:scale result}
\end{table*}

\begin{table}[htbp!]
\vspace{-2mm}
\caption{Ablation Study: The effectiveness of the RMF, stage-by-stage training strategy and stage-wise loss. }
\vspace{-2mm}
\centering
	\resizebox{.48\textwidth}{!}{
\begin{tabular}{|c|c|c|c|}
\hline
            & RMFs & Stage-By-Stage Strategy & Stage-Wise Loss \\ \hline
Fidelity    & $0.6dB \uparrow$                & $0.4dB \uparrow$           & $0.2dB  \uparrow$                      \\ \hline
Flexibility & $\times$                   & $\checkmark$               &  $\checkmark$                           \\ \hline
\end{tabular}}
\label{table:Ablation Study}
\end{table}

\begin{figure}[!htbp]
\centering
\vspace{-1mm}
\includegraphics[width=1\columnwidth]{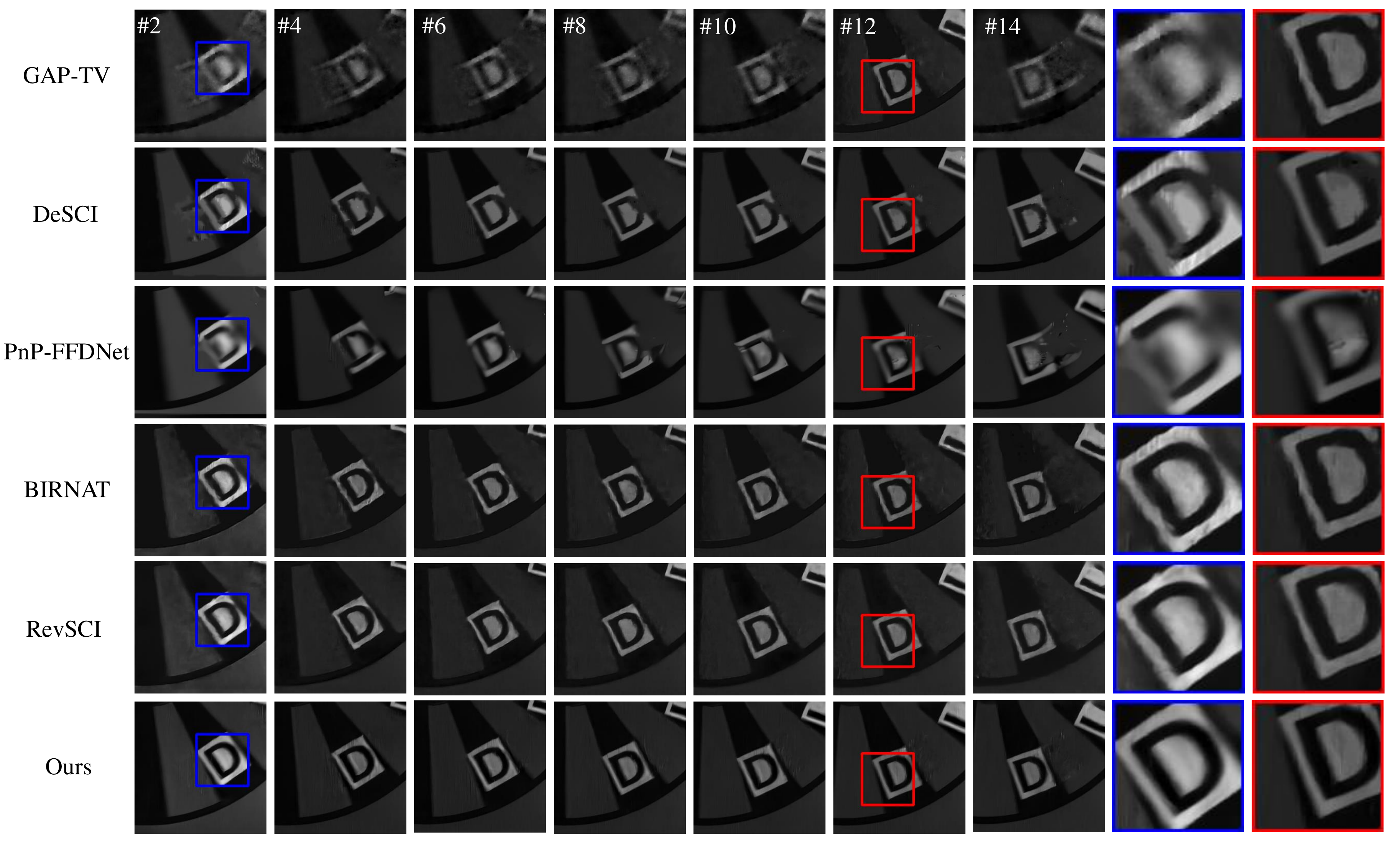}
\vspace{-6mm}
\caption{Comparison of selected reconstruction video frames of real data \textbf{Wheel} with the spatial size of $256\times256\times14$.}
\label{fig:real wheel}
\end{figure}

\subsection{Real Data}
\begin{figure}[!htbp]
\centering
\vspace{-3mm}
\includegraphics[width=1\columnwidth]{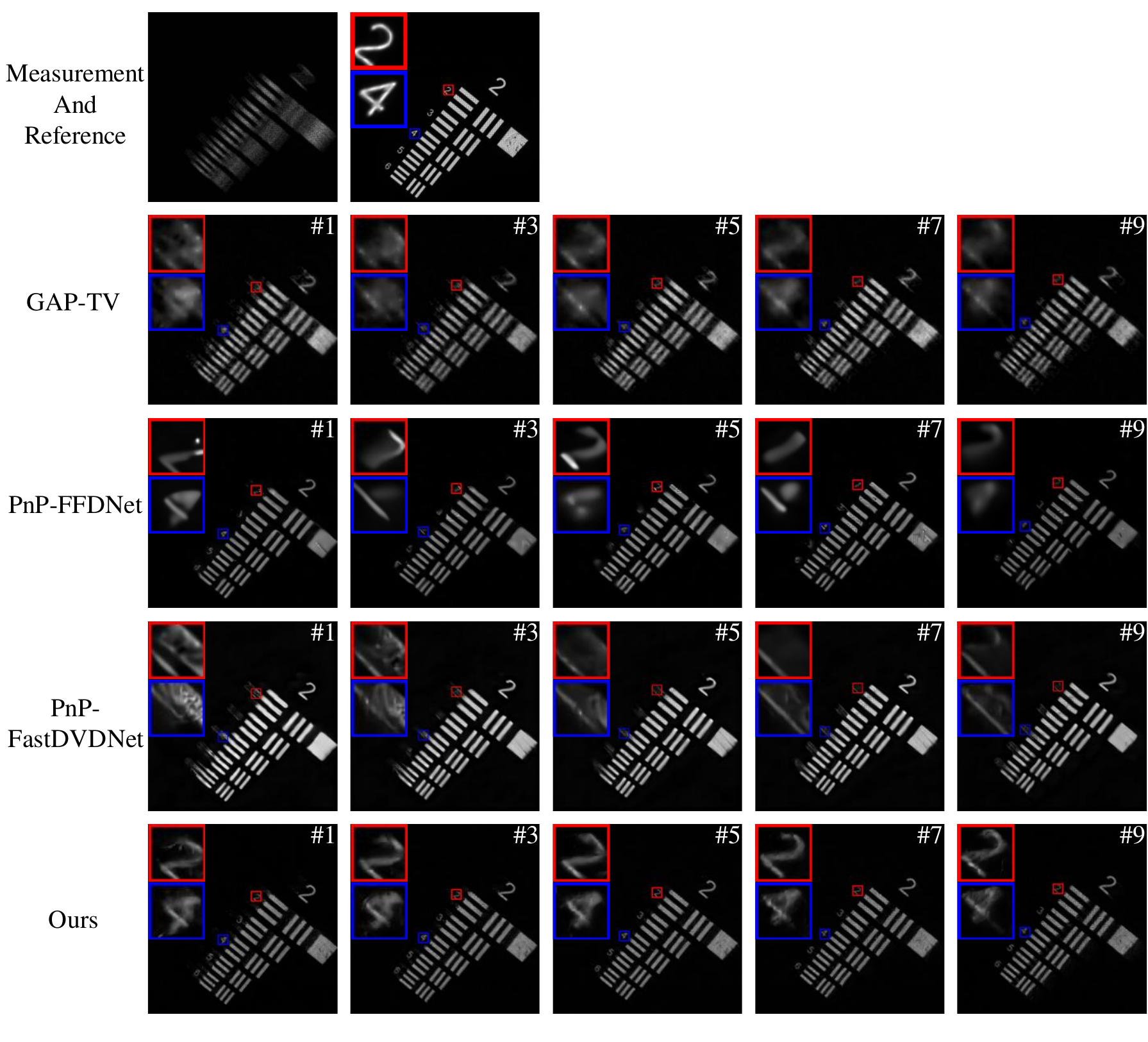}
\vspace{-7mm}
\caption{Comparison of selected reconstruction of real data \textbf{Resolution Chart} with the spatial size of $1000\times1000\times10$.}
\label{fig:Resolution Chart}
\end{figure}

We now apply our model on the real data \textbf{Wheel} with the size of $256\times256\times14$~\cite{Llull13_OE_CACTI}. Reconstructing the real data is harder than simulation data due to the fact that there is unavoidable measurement noise in real data. 
The results of \textbf{Wheel} are shown in Fig.~\ref{fig:real wheel}, where we can see that our proposed model achieves clearer background and sharper edges of the letter `$D$' and the white block than other algorithms. 
To verify the flexibility of our proposed model under real scenarios, we built a new large-scale VCS system using the technique similar to~\cite{qiao2020deep}. The captured measurement, reference and reconstruction results of the dataset \textbf{Resolution Chart} are shown in Fig.~\ref{fig:Resolution Chart}. This is the first time that large scale video frames with the spatial size of `$1000\times1000$' can be reconstructed by the learning-based algorithm due to the memory limitation during training. Note that, thanks to the flexibility of our model, the utilized model can be trained using the data with the spatial size of `$256\times256$'. It can be observed that GAP-TV has severe blurs, PnP-FFDNet and PnP-FastDVDNet are both over smooth, and all of them lose the information of the numbers in the top-left areas. Our model achieves clean results and preserves the detailed information even the patterns on the surface, which further demonstrates the superiority of our model. More real data results are shown in the SM. 
%





\section{Conclusions}
Considering the realistic requirements of fidelity, inference speed and flexibility, we propose a flexible and concise model using minimum stages under the interpretable deep unfolding network. We introduce the invertible structure into the framework to break the bottleneck of memory limitation. Furthermore, we fuse the reference measurement frames into each stage to fully utilize the information from the measurement. We further introduce a stage-by-stage training strategy and utilize stage-wise loss to improve the performance of our model. 
\setlength\parskip{.05\baselineskip}

We used to consider that learning-based methods outperform optimization-based  methods is for the reason that the network learned the information of the mask which is the information of the measurement matrix in general compressive sensing problems. However, the demonstrated flexibility of our proposed model to mask shows that there is a prior better than the conventional hand-crafted priors that can be learned by decoupling the information of the measurement matrix from the network.

\clearpage
\clearpage
\newpage

{\small
\bibliographystyle{ieee_fullname}
\bibliography{main}
}

\end{document}